\documentclass[useAMS,usenatbib,referee]{mn2e}
\usepackage{graphicx,graphics, amsmath}
\title[SWJ2000.6+3210]{Revisiting SWJ2000.6+3210 : A persistent Be X-ray pulsar ? }
\author[P. Pradhan, C. Maitra, B. Paul, B.C. Paul]{Pragati Pradhan$^{1,2}$ Chandreyee Maitra$^{3,4}$
\thanks{E-mail: cmaitra@rri.res.in;} Biswajit Paul$^{3}$ and B.C. Paul$^{2}$\\
$^{1}$ St. Joseph's College, Singamari, Darjeeling-734104, West Bengal, India\\
$^{2}$ North Bengal University, Raja Rammohanpur,  District Darjeeling-734013, West Bengal, India \\
$^{3}$ Raman Research Institute, Sadashivnagar, Bangalore-560080, India\\
$^{4}$ Joint Astronomy Programme, Indian Institute of Science, Bangalore-560012, India\\ }

\begin{document}

\date{}

\maketitle

\label{firstpage}

\begin{abstract}
We present a detailed timing and spectral analysis of the Be X-ray binary 
SWJ2000.6+3210 discovered by the BAT Galactic plane survey. Two \emph{Suzaku} observations of the
source made at six months interval, reveal pulsations at $\sim 890 s$ for both
observations with a much weaker pulse fraction in the second one. Pulsations are 
clearly seen in the energy band of 0.3-10 \rmfamily{keV} of XIS for both observations and at high energies up to 40 \rmfamily{keV}
for the second observation. The broad band X-ray spectrum is consistent with a powerlaw and high
energy cutoff model along with a hot blackbody component. No change in spectral parameters is detected between the observations. We have also analyzed several short observations of the source with \emph{Swift/XRT} 
and detected only a few percent variation in flux around a mean value of $3.5 \times \ensuremath{10^{-11}\,  \mathrm{erg}\,  \mathrm{cm}^{-2}\,\mathrm{s}^{-1}}$. 
The results indicates that SWJ2000.6+3210 is a member of persistent Be X-ray binaries
which have the same broad characteristics as this source. 

\end{abstract}
\section{Introduction}

SWJ2000.6+3210 is a high mass X-ray binary (HMXB) which was discovered during a deep survey with the Burst Alert Telescope (BAT)
instrument onboard the \emph{Swift} observatory \citep{markwardt2005,ajello2008}. It had a time averaged flux (15-55 \rmfamily{keV}) of $2.37 \times 10^{-11}$ erg $cm^{-2}s^{-1}$ \citep{voss2010}.
Optical spectroscopic studies of the source established it as a Be star with a early B V or mid B III spectral type and average magnitude 
of 16.1 (R) \citep{masetti2008}. The same study also put a constraint on the distance to the object at $\sim$ 8 kpc. \\
 SWJ2000.6+3210 was observed twice with \emph{Suzaku} in 2006 with an
interval of of six months with about 10 ks exposures.
 It was also observed several times with the \emph{Swift}/XRT around the same time but only for short intervals.
 Timing and spectral analysis of the \emph{Suzaku} observations was reported by  
\citet {morris2009}. They detected pulsations  
in the second observation with a periodicity of 1056 s and pulsations were not found in the first observation. A pulse phase averaged spectral analysis
was also performed. \\
We have performed a very detailed analysis of the source with both \emph{Suzaku} and 
 \emph{Swift}/XRT pointed observations. The \emph{Suzaku} observations
 have established pulsations of the source and the pulse profiles have been probed in different energy bands, even in
 the hard X-ray band for one of the observations. Detailed spectral analysis has been performed. 
 In section 2, we describe the details of the observations and data reduction. In 
section 3, we describe the timing analysis including pulse period detection and the energy dependent
pulse profiles. In section 4, we present the pulse phase averaged spectral analysis followed by discussions $\&$ 
 in section 5 and conclusions in section 6.

\section[]{Observations \& Data reduction}

\emph{Suzaku} \citep {mitsuda2007} is a broad band X-ray observatory and covers the energy range of 
0.2-600 \rmfamily{keV}. It has two main instruments, the X-ray Imaging Spectrometer XIS \citep {koyama2007}
covering 0.2-12 \rmfamily{keV} range and the Hard X-ray Detector (HXD) having PIN diodes \citep {takahashi2007} which covers and energy range of 10-70 \rmfamily{keV} and GSO crystal scintillators
detectors covering 70-600 \rmfamily{keV}. The XIS consists of four CCD detectors of which three are front illuminated 
(FI) and one is back illuminated (BI).\newline SWJ2000.6+3210 was observed twice with \emph{Suzaku} in 2006. The
observations were seperated by six months and were carried out at the 'HXD nominal' pointing position. 
The XISs were operated in 'standard' data mode in the window off option which gave a 
time resolution of 8 s. The details of the observations are listed in Table $1$ (Henceforth Obs 1. and
Obs 2.).  
The data reduction was done following the Suzaku ABC guide. For the XIS data, all 
unfiltered event files were reprocessed with the CALDB version 20120428 and HEAsoft version 6.11. The source 
being very faint the XIS event files were free from photon pile-up. XIS light curves and spectra
were extracted from the reprocessed XIS data by choosing circular regions of $4^{'}$ radius
 around the source centroid. Background light curves and spectra were extracted by selecting 
 regions of the same size away from the source.
The XIS count rate was 3.8 c/s and 4.9 c/s for Obs 1 and 2 respectively
Response files and effective area files were 
generated by using the 
FTOOLS task 'xisresp'. For extracting  HXD/PIN light curves and 
spectra, unfiltered event files were reprocessed
\footnote{http://heasarc.nasa.gov/docs/suzaku/analysis/hxd
\_repro.html}.
For HXD/PIN background, simulated 'tuned' non X-ray background event 
files 
(NXB) corresponding to the month and year of the respective observations 
were used to estimate the non X-ray 
background \citep{fukazawa2009} \footnote{http://heasarc.nasa.gov/docs/suzaku/analysis/
pinbgd.html}, and the cosmic X-ray 
background was simulated as suggested by the instrument team\footnote
{http://heasarc.nasa.gov/docs/suzaku/analysis/pin\_cxb.html} applying 
appropriate 
normalizations for both cases. Response files of the respective 
observations were obtained from 
the \emph{Suzaku} guest observatory facility.\footnote{http://
heasarc.nasa.gov/docs/heasarc/caldb/suzaku/ and used for the HXD/PIN 
spectrum} \\
\emph{Swift} is a gamma-ray burst explorer \citep{gehrels2004} having three
co-aligned instruments on board, the Burst Alert Telescope (BAT, \cite{barthelmy2005}), 
the X-Ray Telescope (XRT, \cite{burrows2005})
and the Ultraviolet/Optical Telescope (UVOT, \cite{roming2005}). The XRT consists of
 a grazing incidence Wolter I telescope operating at 0.2-10 \rmfamily{keV} energy range. \\
 Nine pointed observations of SWJ2000.6+3210 were made with the \emph{Swift}/XRT starting from
 2005 November to 2006 December. Each pointing is however of short duration, with a
 maximum exposure of 4.5 ks for the last observation. The details of the observations are listed in Table $2$.
The XRT data were reprocessed with  XRTPIPELINE v0.12.6 and were filtered and screened with the
standard criteria. The data in Photon Counting mode was used. Source spectra were extracted 
from circular regions of 20 pixels radius centred on the source position as determined
by the FTOOLS task 'XRTCENTROID'. The background spectra were extracted away from the source with a 
radius of 60 pixels to get a better average value. Ancillary
response files generated with the task 'XRTMKARF' and spectral redistribution
matrix v011 (CALDB ver 20120209). 
 
 \begin{table} 
\caption{List of \emph{Suzaku} observations }
\label{table:1}
\centering
\begin{tabular}{c  c   c}
\hline \hline
Time & Observation Id & Effective exposure (ks) \\
\hline 
2006-04-12 & 401053010 (Obs. 1)& 9.9  \\
2006-10-31 & 401053020 (Obs. 2)& 11.7 \\
\hline
\end{tabular}
\end{table}

 \begin{table} 
\caption{List of \emph{Swift} observations }
\label{table:2}
\centering
\begin{tabular}{c  c   c}
\hline \hline
Time & Observation Id & Effective exposure (ks)\\
\hline 
2005-11-09 & 00035237001 (Sw. 1)&  1.23  \\
2006-10-13 & 00035237004 (Sw. 2)&  1.90 \\
2006-11-17 & 00035237005 (Sw. 3)&  2.29  \\
2006-11-25 & 00035237007 (Sw. 4)&  1.16  \\
2006-12-08 & 00035237009 (Sw. 5)&  4.49  \\

\hline
\end{tabular}
\end{table}

\section{timing analysis}
Timing analysis was performed on the XIS and PIN data of the \emph{Suzaku}
observations after applying barycentric corrections to the 
event data files using the FTOOLS task 'aebarycen'. \emph{Swift}/XRT observations
had short exposures and hence did not have enough sensitivity to search for pulsations
or perform any further timing analysis. Light curves with a time resolution of 8 s (full window mode of the XIS data)
and 1 s were extracted from the XISs (0.2--12 \rmfamily{keV}) and PIN (10--70 \rmfamily{keV}) respectively. Timing analysis was performed
by summing the XIS 0, 1, 2 and 3 light curves after background subtraction. Figure 1 shows the XIS and PIN
light curves of Obs. 1 and Obs. 2 with a binsize of 8 secs.

\subsection{Pulse period determination}

Initially pulsations were searched in both the observations by applying the epoch folding technique using the FTOOLS 'efsearch'. Pulsations were weaker in Obs.1  but 
were clearly detected in both the observations with periods of  889.7 $\pm$  $4.7 $ s  
\& 887.6 $\pm$  $2.8$ s for Obs. 1 and 2 respectively. The pulse period determined is 
consistent for both the observations, and is in contradiction with that found by \cite{morris2009} who claimed no pulsations in
the data of Obs. 1 and a period of 1056 s for Obs. 2. 
 
We note that with 'efsearch' a peak in the $\chi^{2}$ is also seen at  $\sim$ 1056 s in the second observation 
which is less
significant than the peak at $\sim 890 s$. Folding the data with this period also reproduced the pulse
profiles given in \cite{morris2009}. To further confirm the correct pulse period of the source, we created
Power Density Spectrum (PDS) using the FTOOL task '\emph{powspec}' from the XIS light 
curves (0.2--12 \rmfamily{keV}, see middle panel of Figure 2). The PDS on both the observations clearly 
shows the peak corresponding to the pulse period determined in this work. Lastly we created Lomb Scargle Periodograms
from the background subtracted and detrended (A best fit linear function subtracted from the light curve)
XIS light curves of the observations (see Figure 3). The periodograms
show clear peaks at the same period. The peaks corresponding to 1056 s are much smaller in amplitude.
All of these establish that pulsations of $\sim$ 890 s exist in the data
which corresponds to the spin period of the neutron star.
Pulse period could not be independently determined from the HXD/PIN data for the first observation (Obs. 1).
Epoch folding performed on PIN data of Obs. 2 however shows the evidence of a peak at the frequency corresponding
to the pulse period. \\

We have further investigated the 1056 s periodicity reported earlier by \cite{morris2009}.
 In Figure 3, we have put markers at 1056 s intervals on the XIS lightcurve. It can be seen that some of the minima
 coincide with the markers while the others show a secular offset. We infer that the minimum actually occur at  $\sim$ 890 s 
intervals, which is the pulse period of the neutron star.
The small peaks seen on either sides of the main peak in the top right panel of Figure 2 are 
aliases due to a observation window that is periodic at the  $\sim$ 96 minutes orbital period of the satellite
satisfying the relation
\begin{equation*}
 P_{alias}={(\frac{1}{P_{spin}}\pm \frac{1}{P_{orb}})}^{-1}
\end{equation*}

\subsection{Pulse profiles and energy dependence}

We created the average low energy pulse profiles from the XIS data (0.3-12 \rmfamily{keV}) by folding the background subtracted 
light curves with the 
respective pulse periods obtained. As is evident from section 3.1, pulsations are weaker in Obs. 1 which are
also reflected in the corresponding pulse profiles, with Obs. 2 having a higher pulsed fraction. We also obtained the hard X-ray pulse profile (10-40 
\rmfamily{keV}) by
folding the background subtracted PIN light curve of Obs. 2 with the pulse period obtained from the XIS data. This clearly shows that hard
X-ray pulsations exist in the source at least upto 40 \rmfamily{keV}. The pulse profiles 
for both the observations are shown in figure 5 for both the XIS and PIN data. As is evident from the figure, the XIS pulse profiles
are complex  with the presence of dips.  Obs. 2 has a 
higher pulse fraction ($\frac{F_{max}-F_{min}}{F_{max}+F_{min}}$) and the pulse fraction from the two observations are 20 \% and 41 \%.
The pulsed fraction in the pulse profile of obs. 2 given in \cite{morris2009} at the reported period of
1056 s is lower $\sim 20 \%$.
 Overall, the PIN pulse
profile also looks similar to that obtained from XIS, with a shift in phase of $\sim$ 0.1 in the minima.  \\
The pulse profiles obtained for the source are typical of that found in many X-ray pulsars, some of which also
exhibit strong energy dependence specially in the range below 10 \rmfamily{keV}. To probe this further, we created
energy resolved pulse profiles with the XIS data (0.3-12 \rmfamily{keV}) for both the observations by folding the light curves in different energy bands 
with the respective pulse periods. Figure 5 shows the energy dependent pulse profiles for Obs. 1 and 2. 
Both the observations show a complex energy dependent nature of the pulse profiles.
The narrow dip at $\sim$ 0.55 for Obs. 1 and $\sim$ 0.9 for Obs. 2 becomes shallower
with energy and disappears at $\sim$ 5 keV. Energy dependence of the pulse profile is complex as 
found in many other X-ray binary pulsars \citep{tsygankov2007,devasia2011,maitra2012}.

\begin{figure}
\includegraphics[scale=0.3,angle=-90]{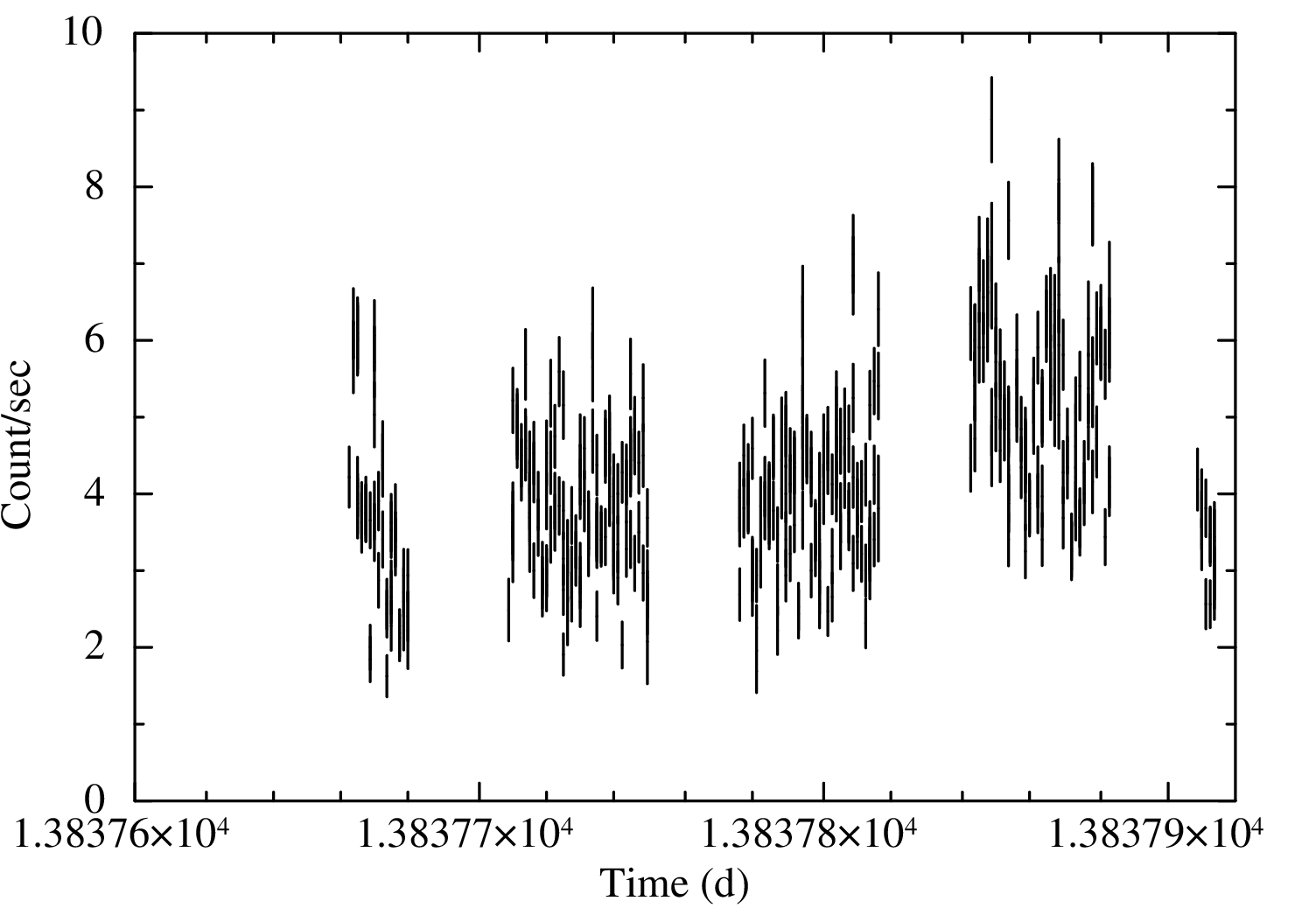}
\includegraphics[scale=0.3,angle=-90]{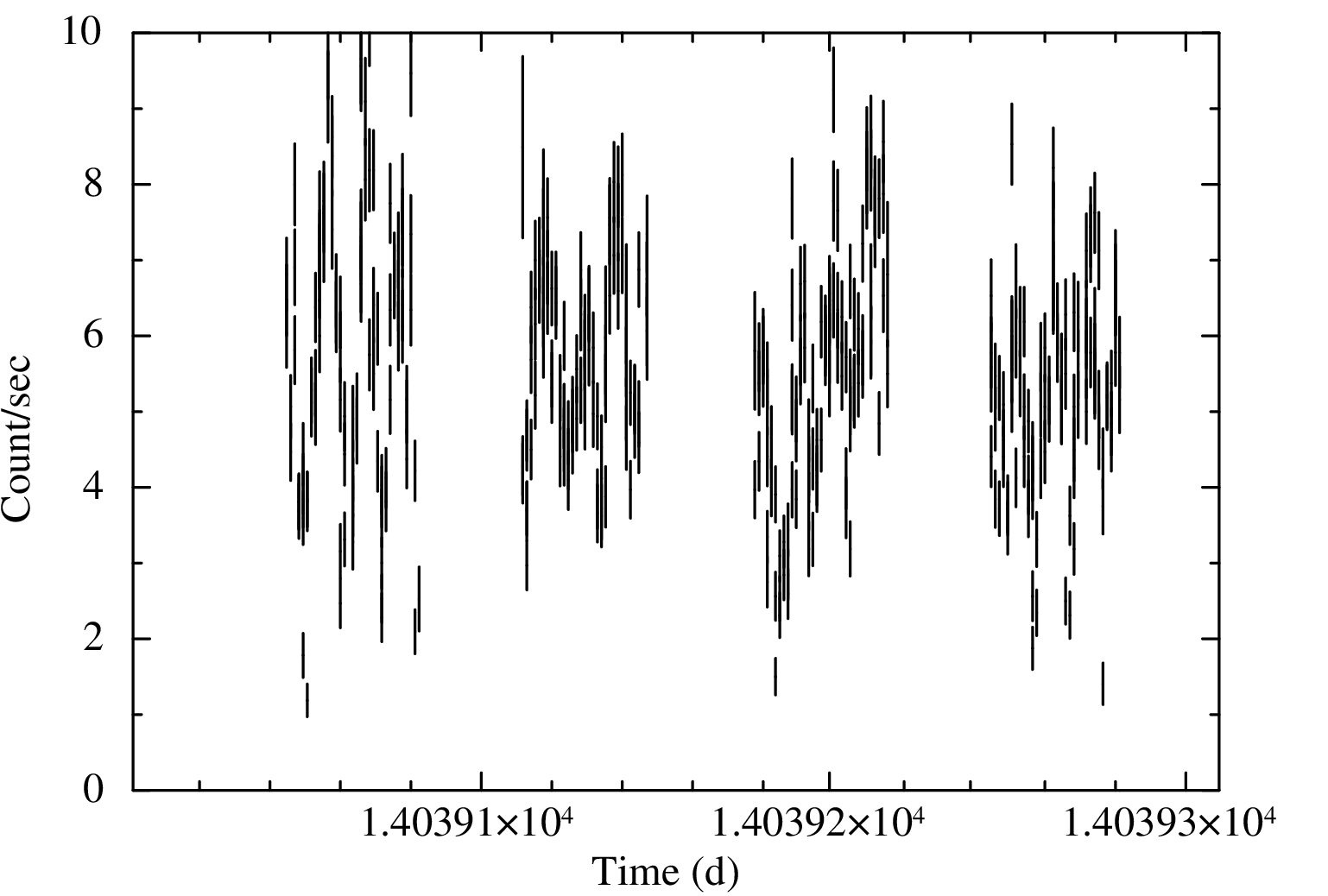}
\caption{The left and right panels show the summed  \emph{Suzaku}/XIS light Curves (0.3-12 \rmfamily{keV}) 
for Obs. 1 and 2 respectively.}
\label{fig1}
\end{figure}

\begin{figure}
\centering$
\begin{array}{cc}
\includegraphics[width=5cm,angle=-90]{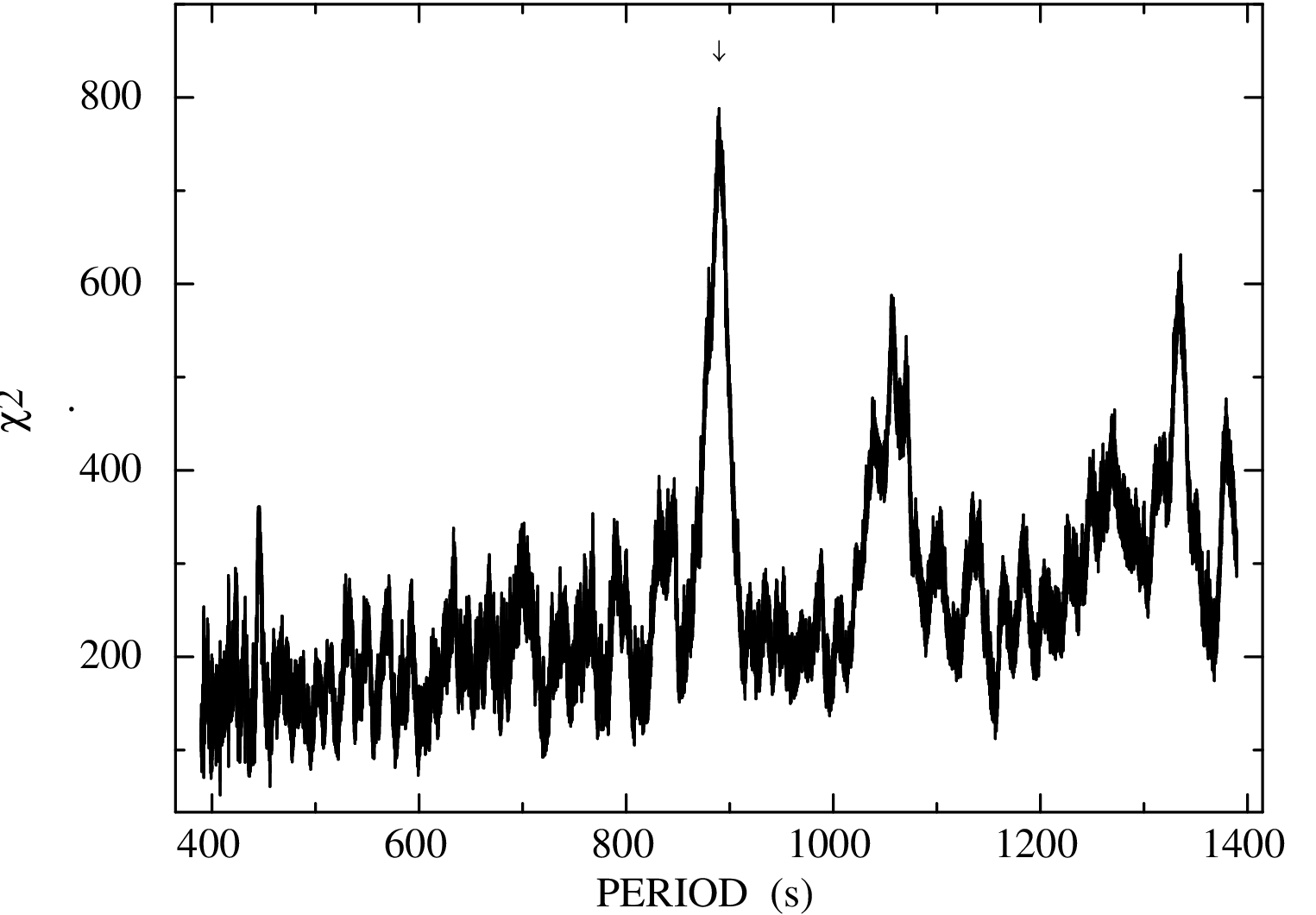} &
\includegraphics[height=7cm,angle=-90]{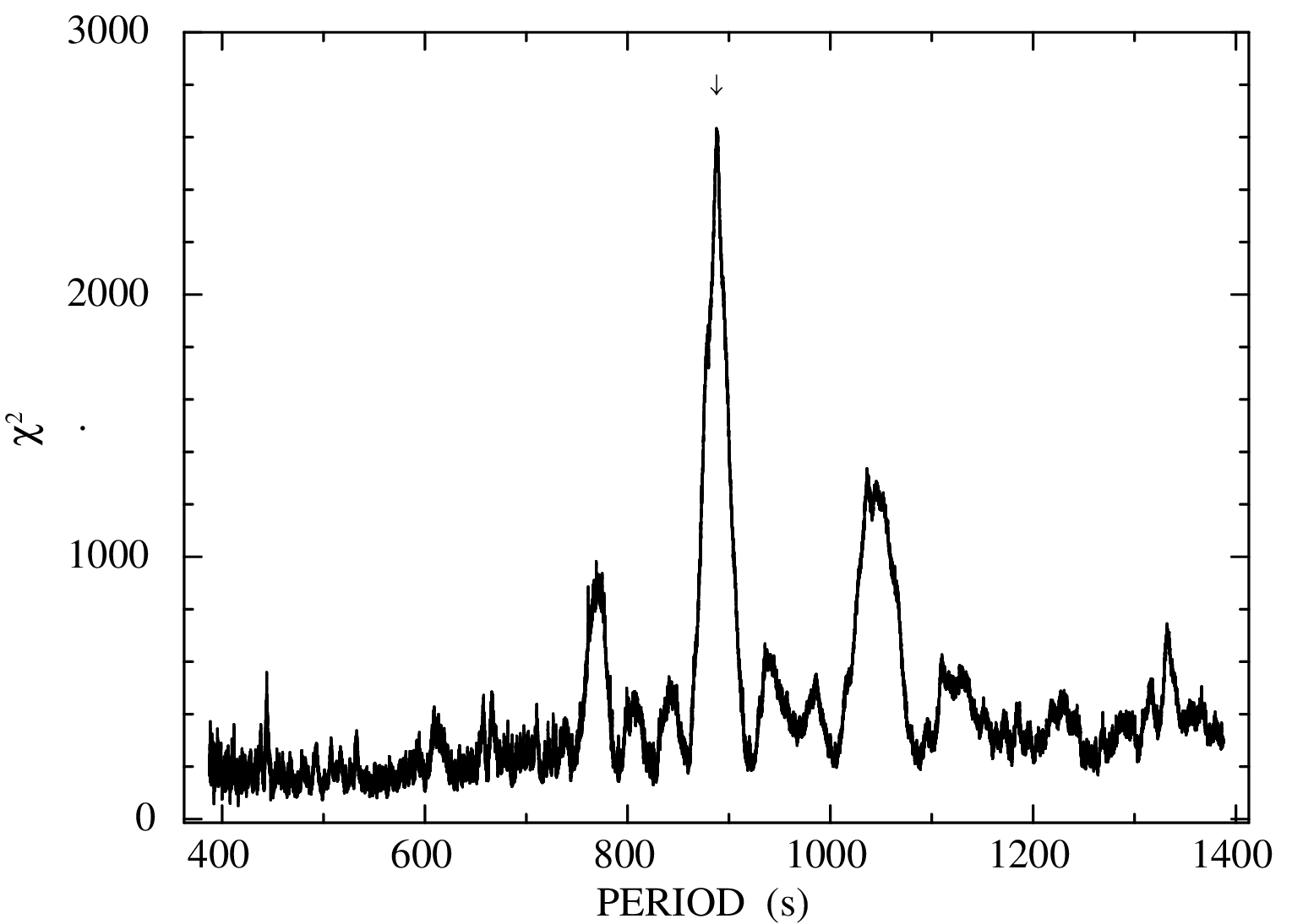} \\
\includegraphics[height=7cm,angle=-90]{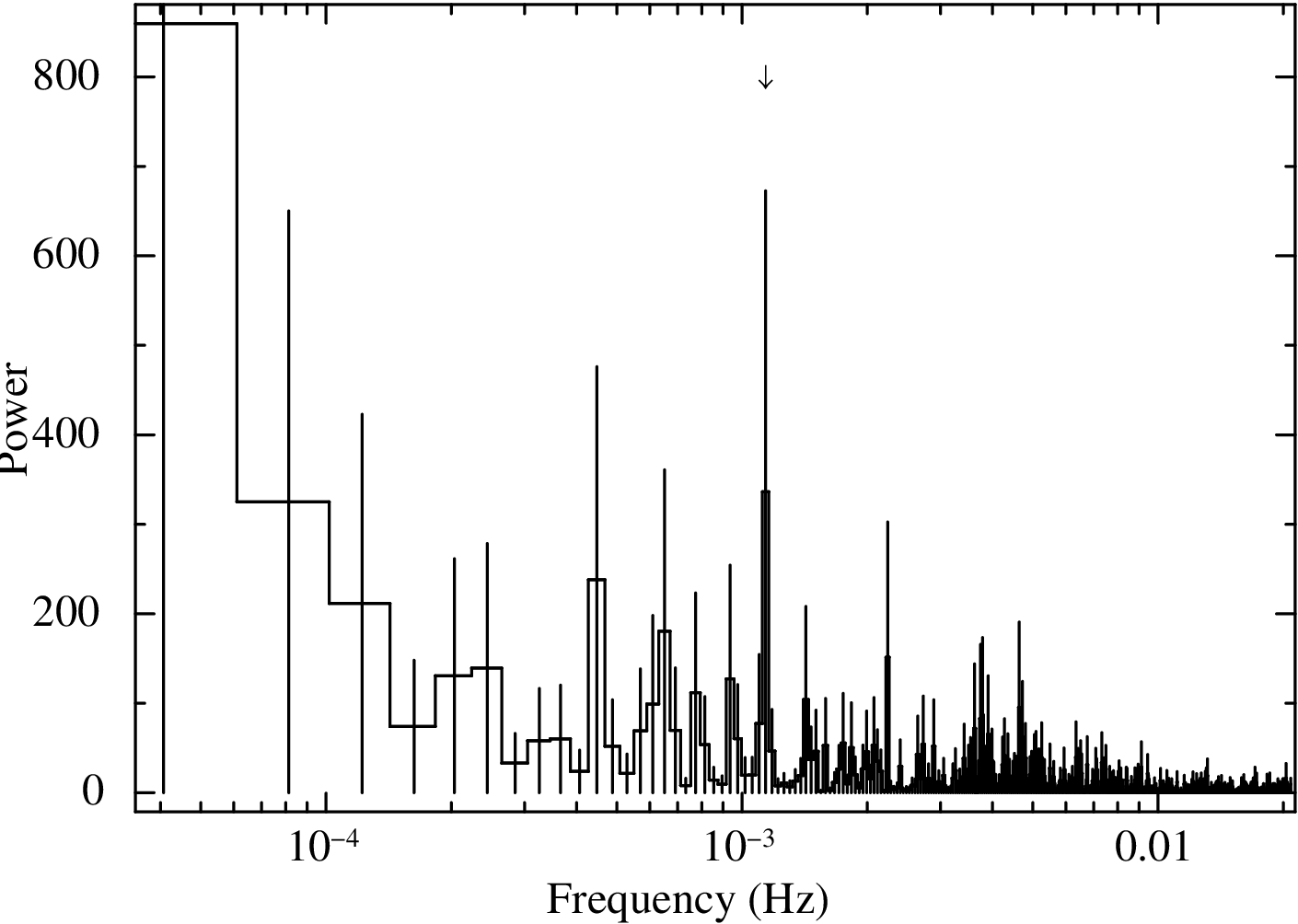} &
\includegraphics[height=7cm,angle=-90]{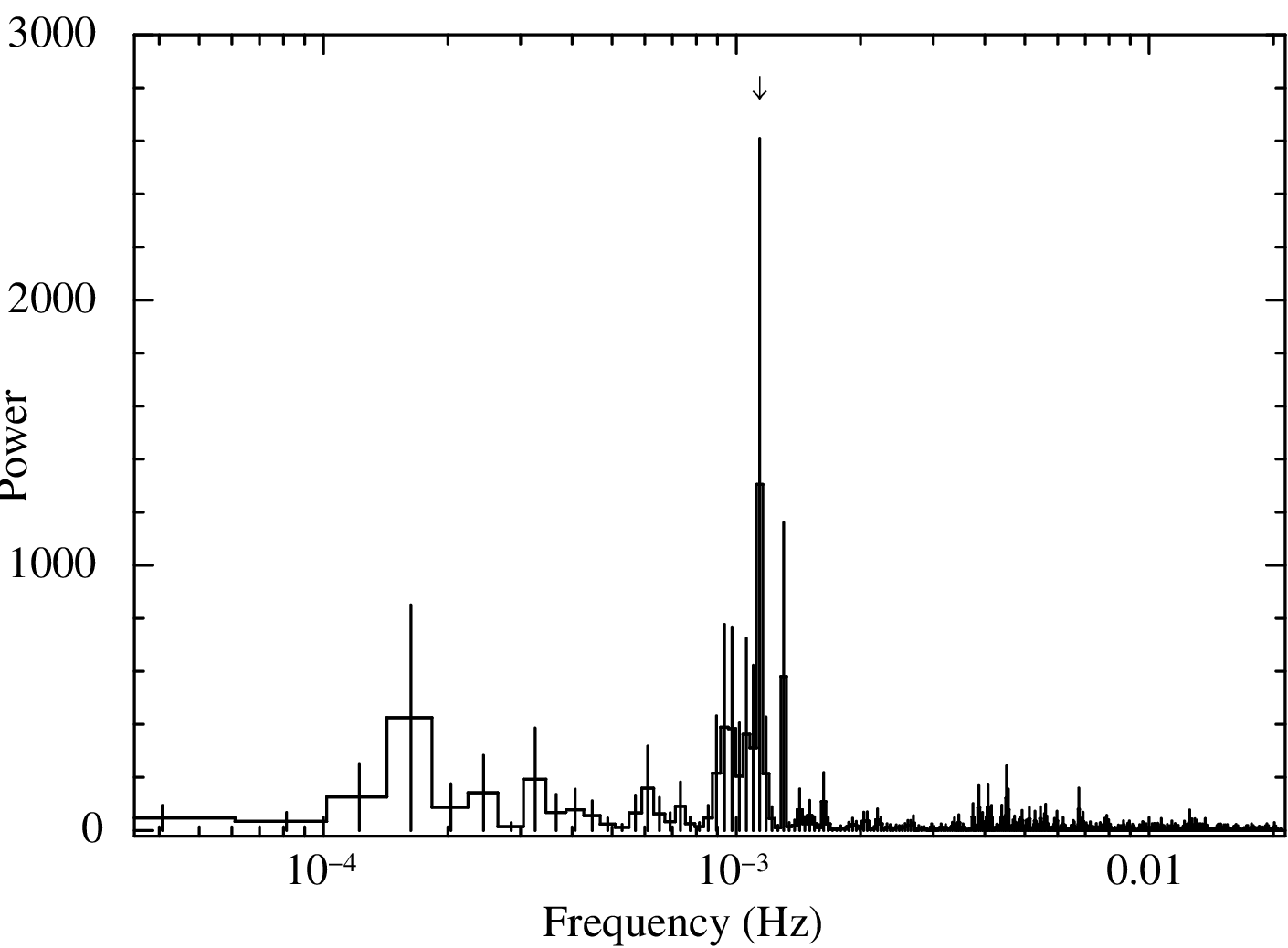} \\
\includegraphics[height=7.6cm,angle=-90]{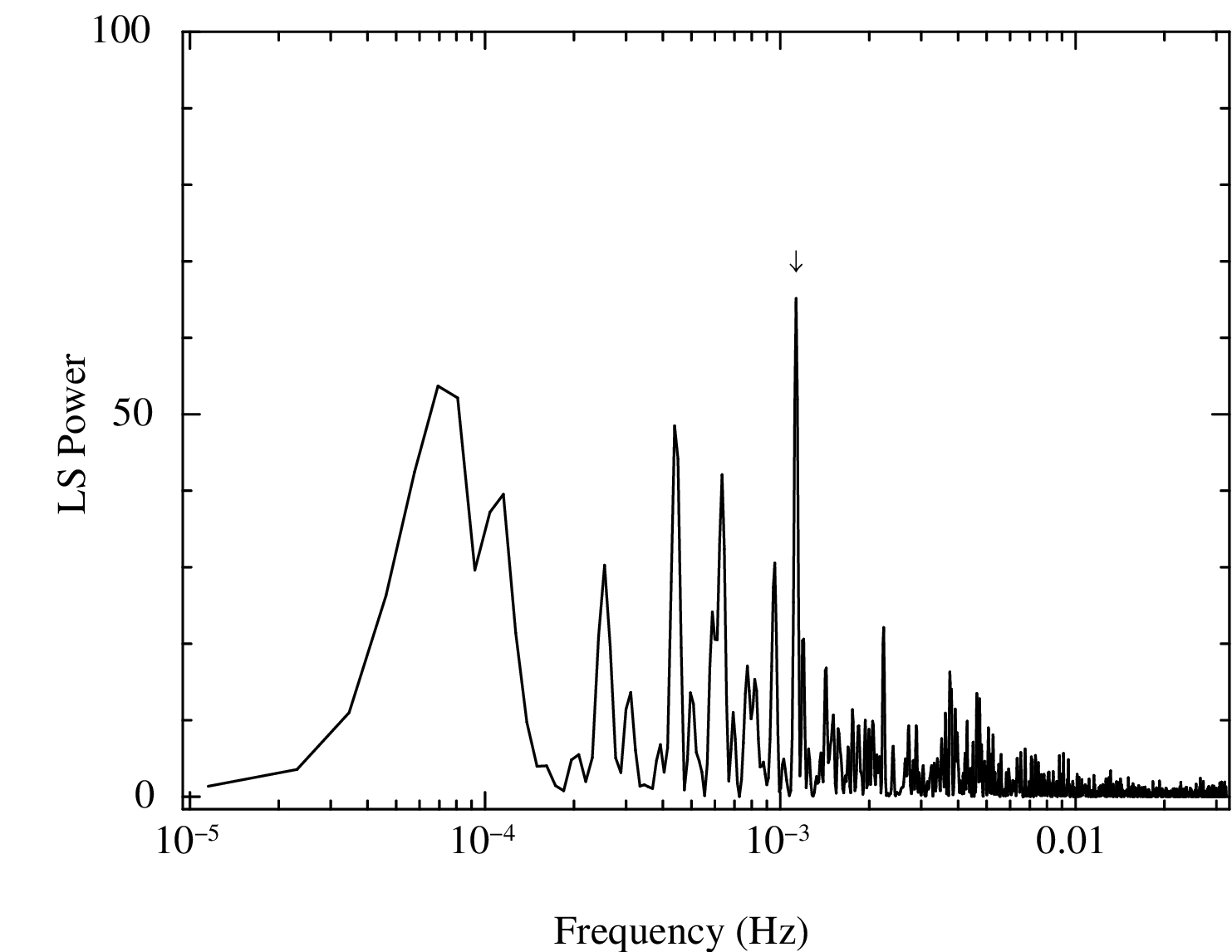} &
\includegraphics[height=7cm,angle=-90]{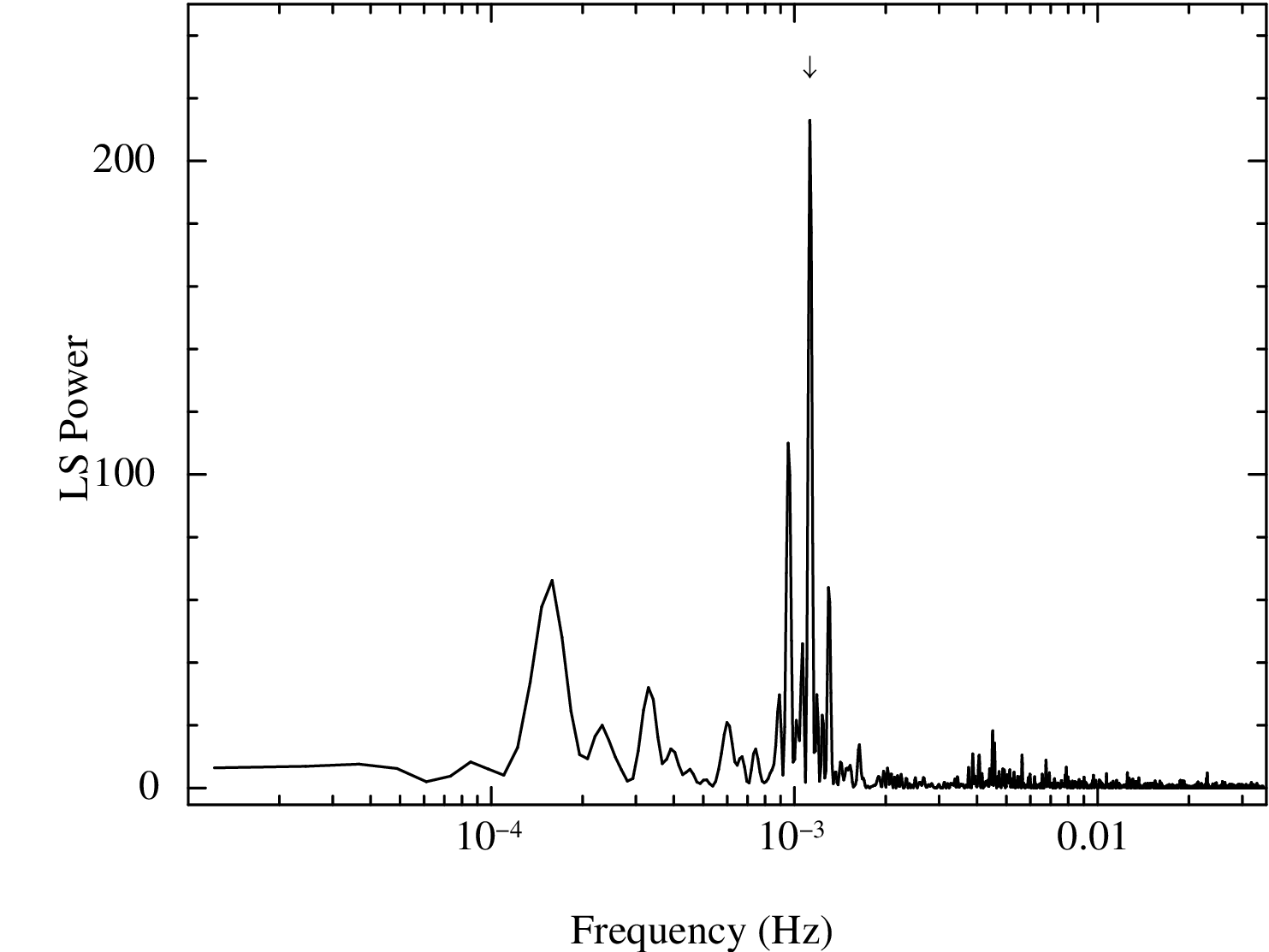} \\
\end{array}$
 \caption{The first panel shows the epoch folding results for \emph{Suzaku} Obs. 1 and 2 (left and right panel) using the
 summed XIS light curves (0.3-12 \rmfamily{keV}). The second panel shows the PDS created for the same 
 light curves showing 
 the frequency corresponding to the pulsation period of the source as indicated
 by the arrows. The third  panel shows the Lomb Scargle Periodogram of the same light curves clearly
 showing the pulsating frequency of the source indicated by arrows.}

\label{fig2}
\end{figure}


\begin{figure}
\begin{center}
\includegraphics[height=18cm,width=4cm,angle=-90]{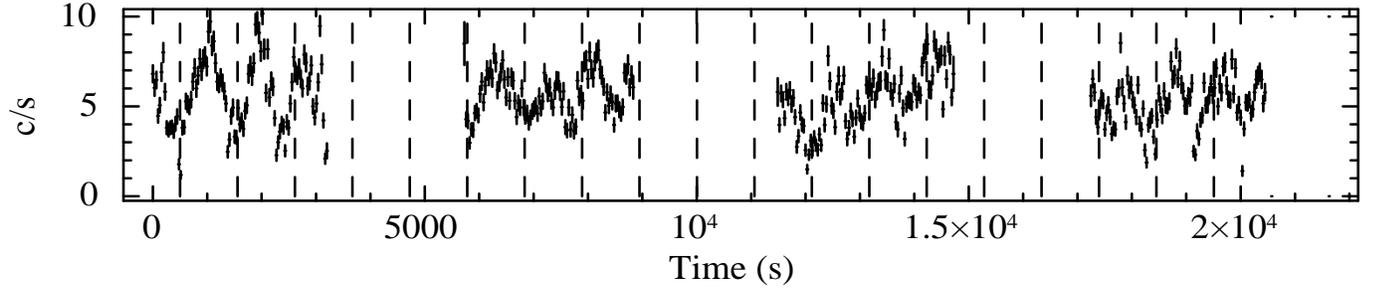}
 \caption{Summed XIS light curve (0.3-12 \rmfamily{keV}) of \emph{Suzaku}
 Obs. 2 marked at every 1056 s, showing the shift in the minima of the light curve
 and return every  5-6 cycle.}
\end{center}
\label{fig4}
\end{figure}

\begin{figure}
\centering
\includegraphics[height=7 cm]{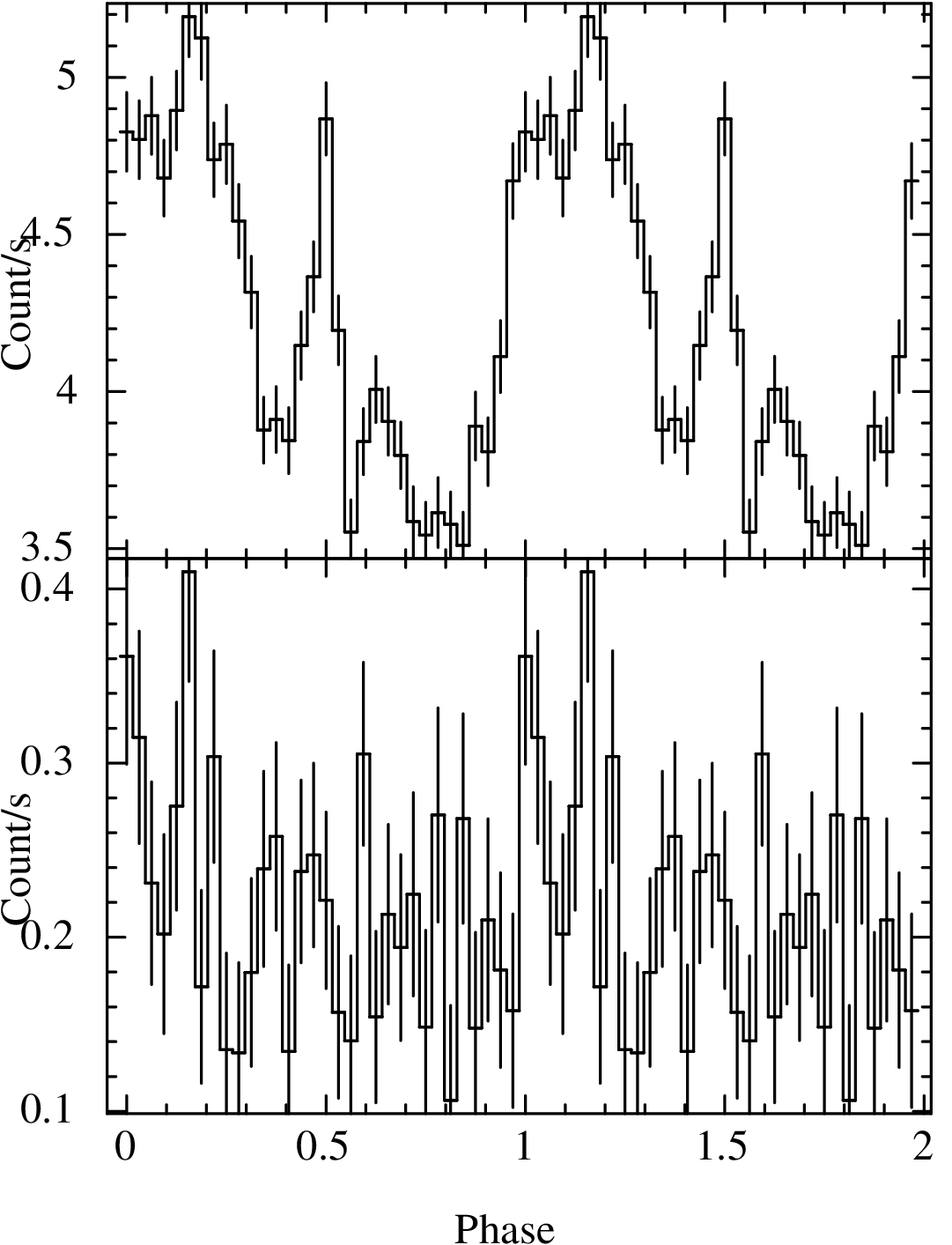}
\includegraphics[height=7 cm]{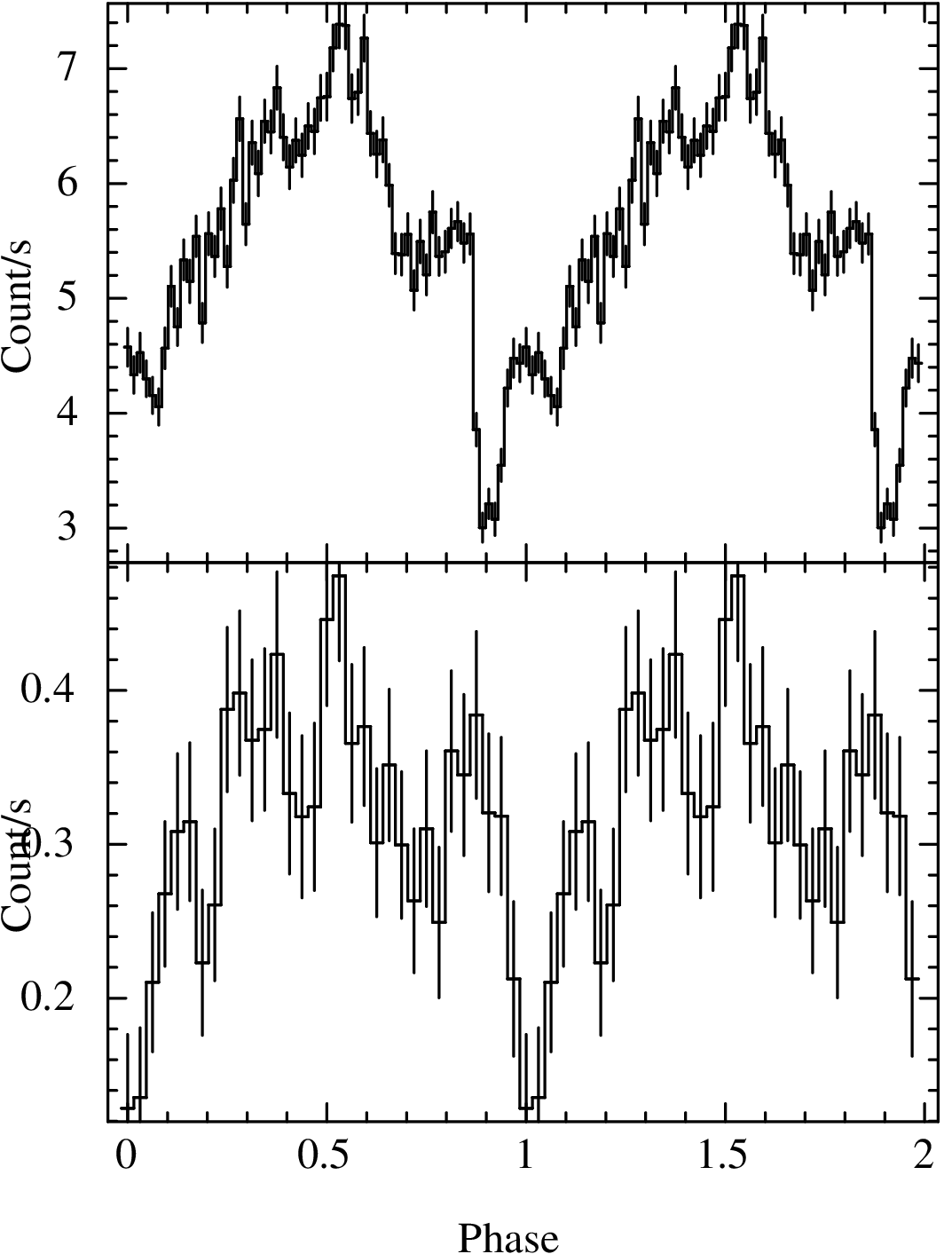}
\caption{The XIS (0.3-12 keV) pulse profiles of the two Suzaku observations are shown in the top 
two panels (left -- Obs. 1 and right -- Obs. 2). PIN profiles of the same observations are shown in the bottom panels.}
\label{fig5}
\end{figure}

\begin{figure}
\centering
\includegraphics[height=8 cm,width=5cm]{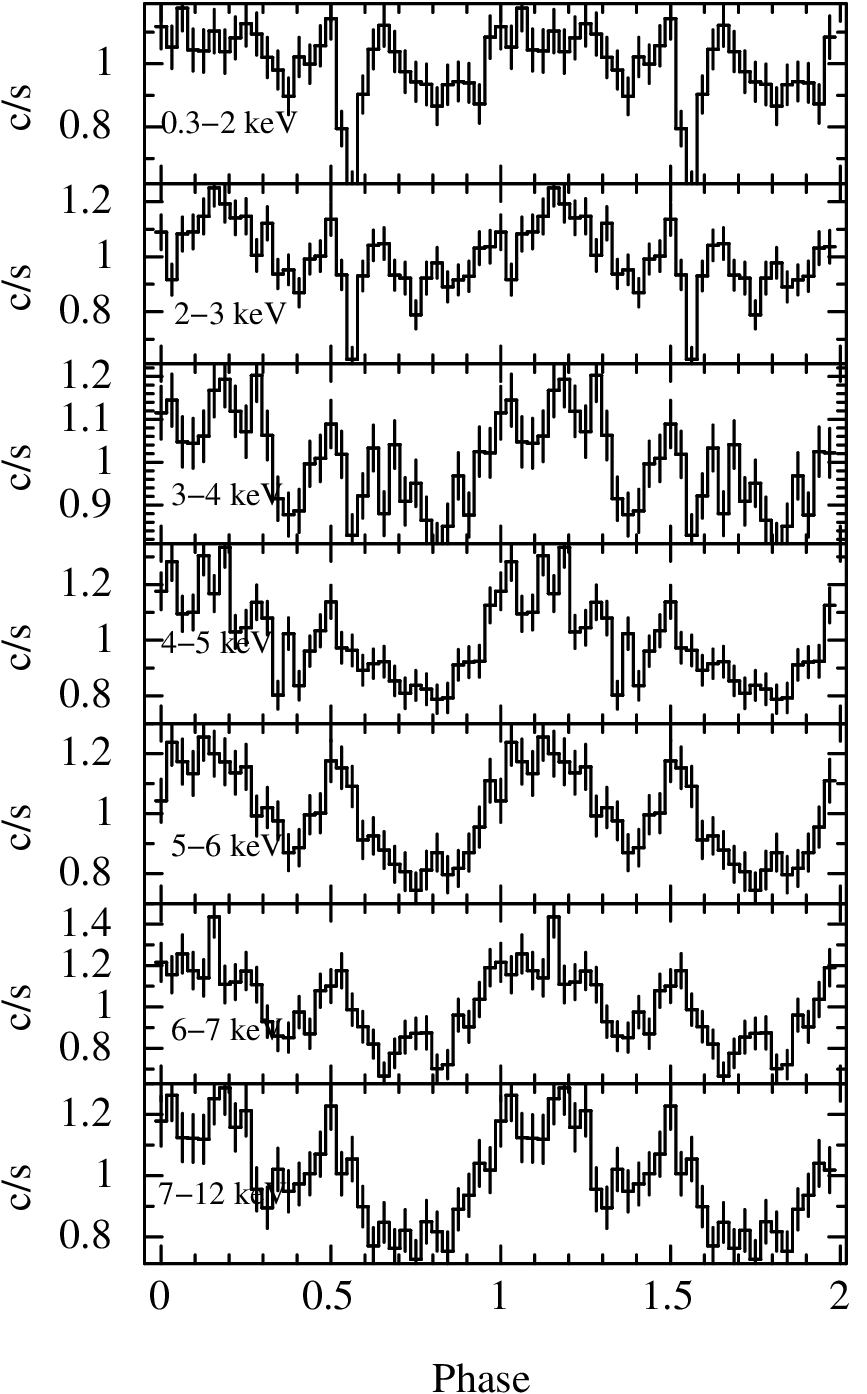}
\includegraphics[height=8 cm,width=5cm]{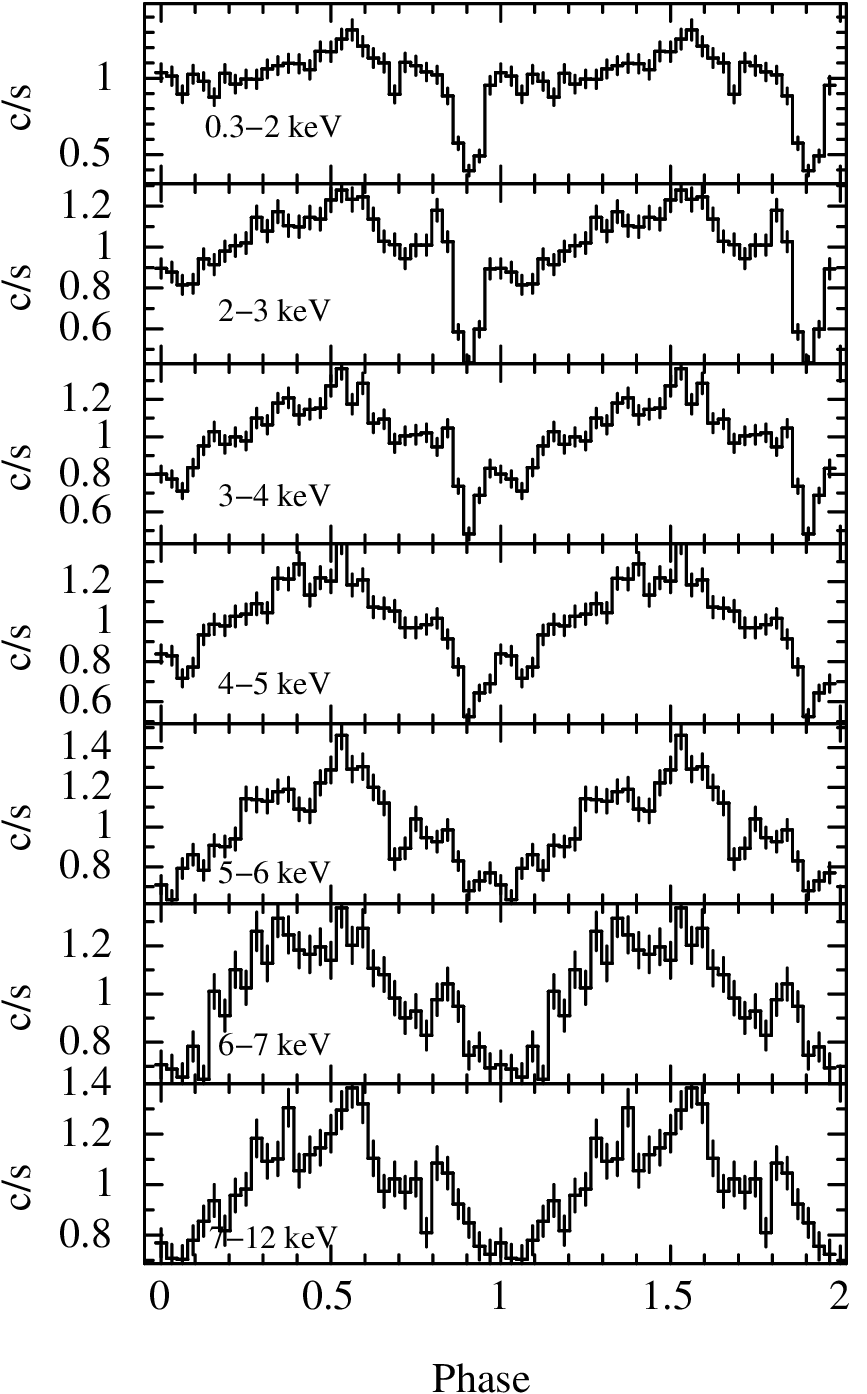}

\caption{The left and right figures show the energy dependent pulse profiles from 0.3-12 \rmfamily{keV}
created using the summed XIS light curves for Obs. 1 and 2 respectively. The pulse profiles denote normalized intensity.}
\label{fig6}
\end{figure}


\begin{figure}
\centering$
\begin{array}{cc}
\includegraphics[height=7cm,angle=-90]{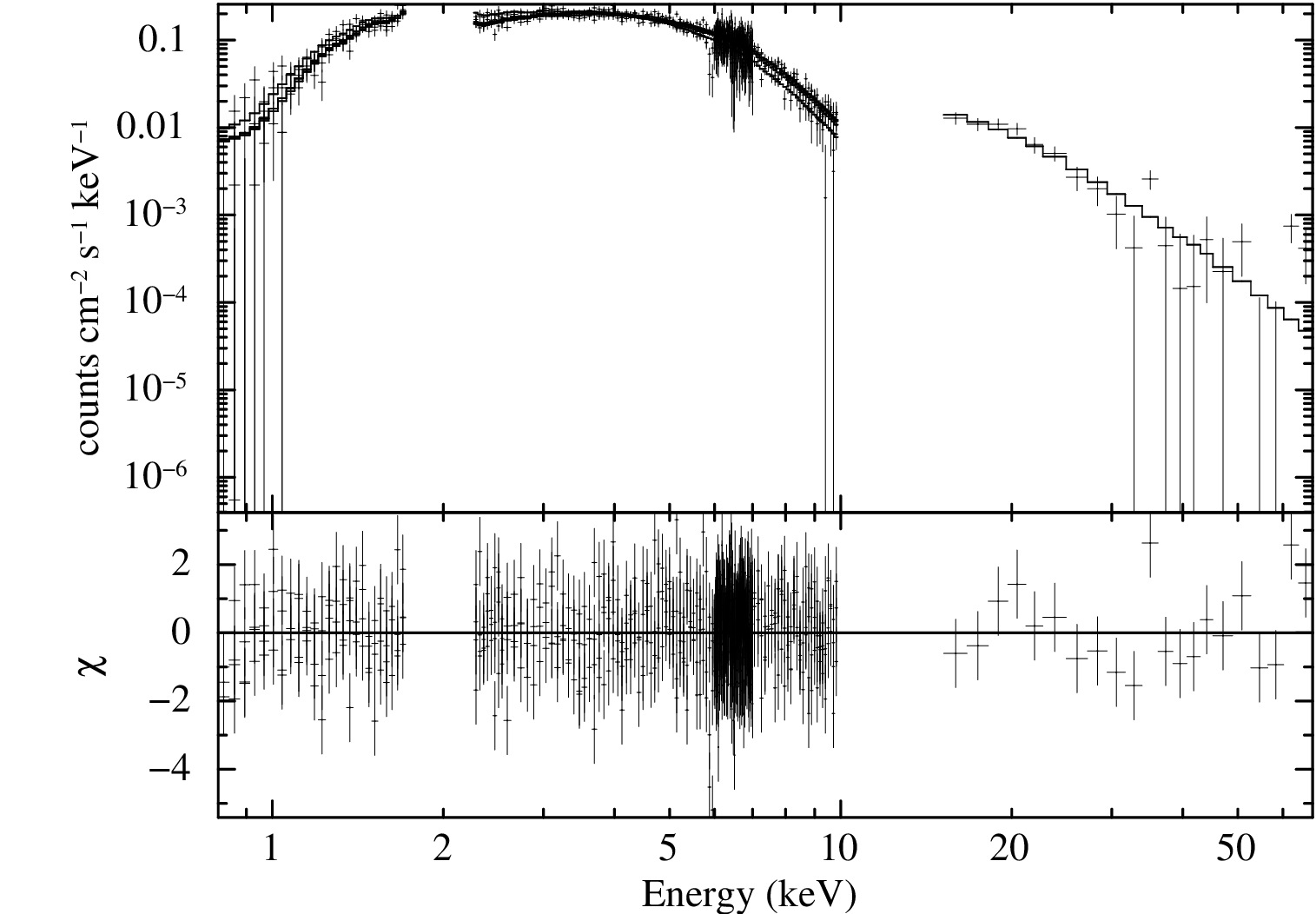} &
\includegraphics[height=7cm,angle=-90]{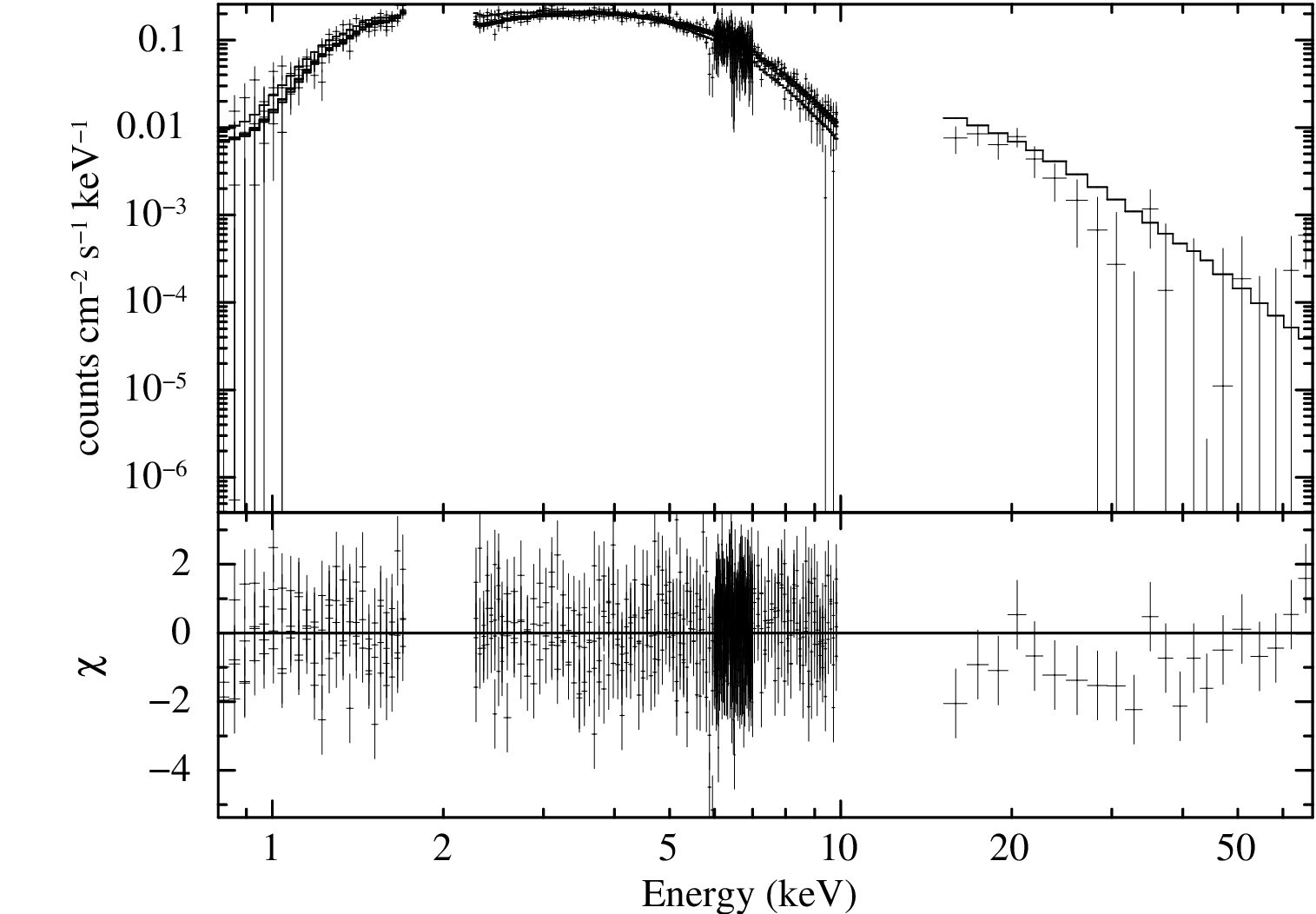} \\
\includegraphics[height=7cm,angle=-90]{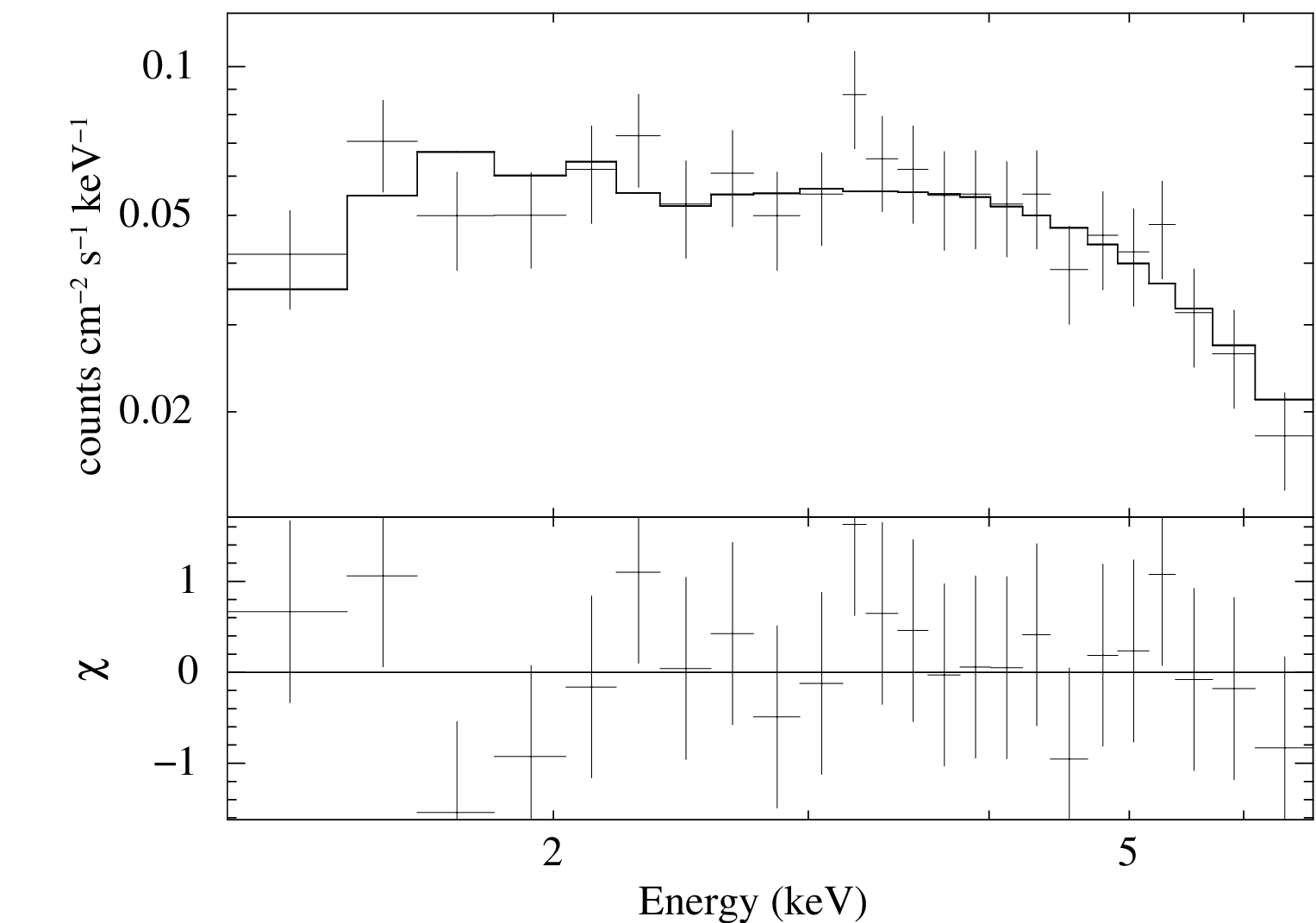} & 
\includegraphics[height=7cm,angle=-90]{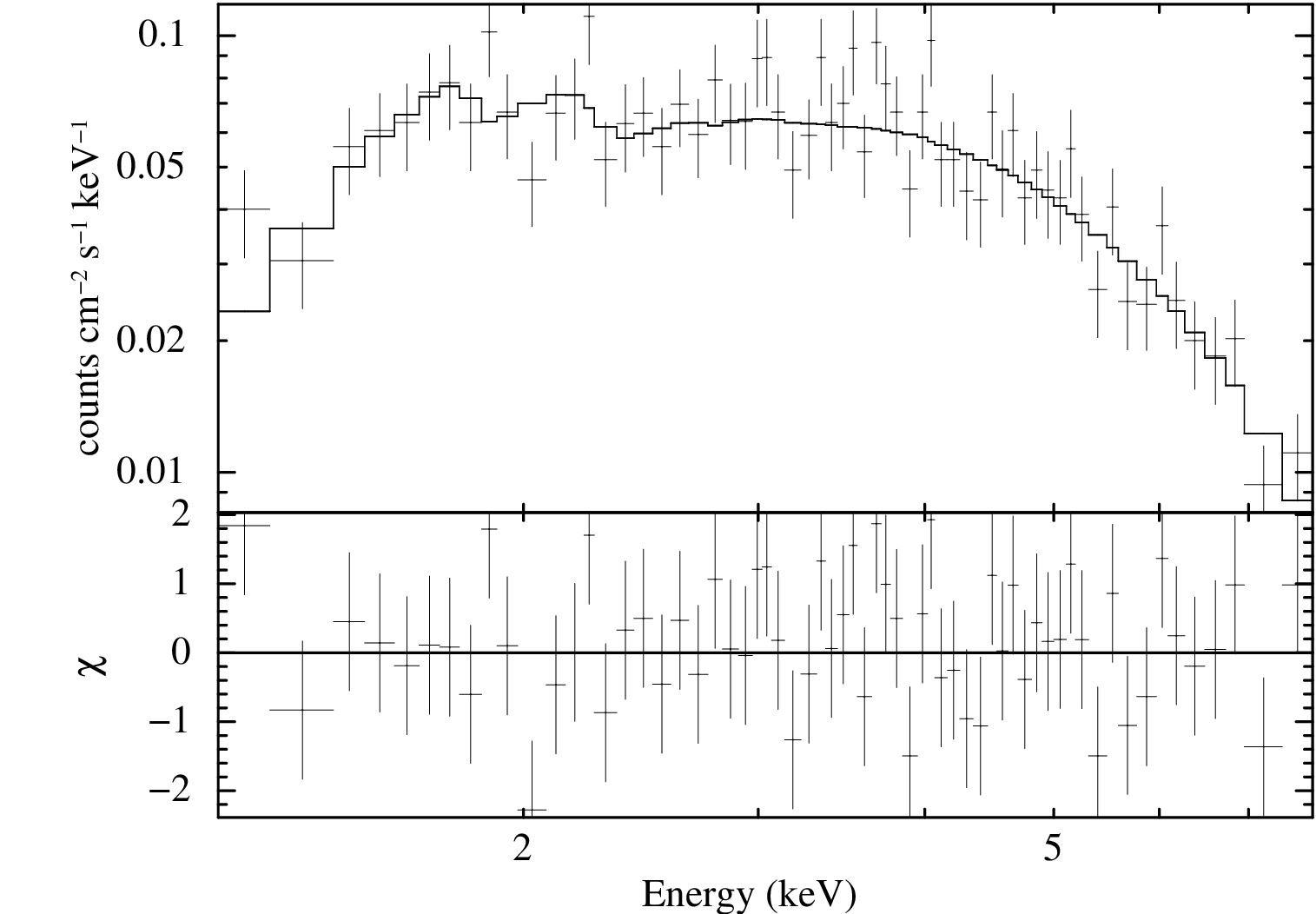} \\
\end{array}$
\caption{The left and right figures of the upper panel show the best fitted phase averaged
energy spectra along with their residuals for \emph{Suzaku} observation 1 and 2 respectively. The XRT spectra
for the same data corresponding to the two longest observations (Swift Obs. Id 00035237004 and Swift Obs. Id 00035237009 respectively) are shown in the left and right figures of the bottom panel.}
\label{fig8}
\end{figure}
\section{spectral analysis}

\subsection{Pulse phase averaged spectroscopy} 
We performed pulse phase averaged spectral analysis 
of SWJ2000.6+3210 using both the \emph{Suzaku} observations
and the two longest observations of XRT (Obs. Id 00035237004 and 00035237009). 
The other available observations of XRT were used only to estimate the flux of the source
 on that day and were not used for spectral parameter estimations due to
their very short exposures.\\
For \emph{Suzaku}, we used 
spectra from the front illuminated CCDs
(XISs 0, 2 and 3), back illuminated CCD (XIS-1) and the
 PIN.
Spectral fitting was performed using \emph{XSPEC} v12.7.0. The energy 
ranges chosen for the fits were 0.8-10 \rmfamily{keV}
for the XISs and 15-70 \rmfamily{keV} for the PIN spectrum 
respectively. Due to an artificial structure in the XIS spectra around 
the Si edge and Au edge, the  energy range of 1.75-2.23 \rmfamily{keV}
was  
neglected. After appropriate background subtraction we fitted the 
spectra simultaneously with all parameters tied, except the relative 
instrument normalizations 
which were kept free. The XIS spectra were rebinned by a factor of 16 throughout the 
energy range. 
The PIN spectra were rebinned by a factor of 4 upto 22 \rmfamily{keV}, by 6
upto 45 \rmfamily{keV}, and 10 upto 70 \rmfamily{keV}. \\
For \emph{Swift}/XRT, the spectra were grouped to a minimum of 20
counts per bin and fitted in the energy range of 0.8-8 \rmfamily{keV}. \\
At first the \emph{Suzaku} spectra were fitted with a single powerlaw and an interstellar
absorption component. The fits however proved to be unsatisfactory with large systematic residuals left after
the fitting and high values of $\chi^{2}_{\nu}$. Addition of a high energy cutoff to the spectra of Obs. 1
and a blackbody component to the spectra of Obs. 2 improved the fits considerably with  $\chi^{2}_{\nu}$
values of $\sim$ 1. The analytical form of the high energy cutoff model as used in \emph{XSPEC} is as follows \\
\[
    I(E)= 
\begin{cases}
  E^{-\Gamma} ,& \text{if } E < Ec\\
   E^{-\Gamma}\exp({\frac{Ec-E}{Ef})} ,              & \text{if} \, E \ge Ec\\
\end{cases}
\]

The best fits were finally obtained with the addition of both the high energy cutoff and the blackbody
component to the absorbed powerlaw model with a $\Delta\chi^{2}$ of 4 and 10 and $\chi^{2}_{\nu}$ values of 1.18 and 1.16 
(343 and 358 d.o.f) for the two observations respectively. The Fe K$\alpha$
line detected by \cite{morris2009} was not found to be significant by us. We derived the upper limits
for a narrow Fe K$\alpha$ line in the spectra by fixing its energy and width at 6.4 and 0.01 keV respectively, 
\textbf{and deriving an upper limit on the normalisation and hence the equivalent width of the line.
The 90\% confidence level upper limits of the equivalent width for a narrow Fe line at 6.4 keV
were derived to be 15 eV and 13 eV for Obs.1 and Obs. 2 respectively.}
The blackbody temperature were $\sim$ 1.19 and 1.50 keV and the radius of the emitting region were $\sim$ $377 \pm 65$ and $345 \pm
58$ m of 8 Kpc and
the same observations respectively. The high
temperature of the blackbody component and the small emission region support its origin from the polar
cap region rather than being a reprocessed component \citep{paul2002}. This aspect is further discussed in 
section 5.\\
 We did not further fit the data with any other
spectral models in \emph{ XSPEC} like 'NPEX', 'FDCUT' or 'CompTT' usually used to fit the HMXBs spectra,
due to the low count rates of the source and hence statistical limitations. \\
Although the source exhibits complex energy dependent pulse profiles, the statistical quality of the data was not sufficient
to perform pulse phase resolved
spectroscopy and further probe the energy dependent nature of the dips. \\
The \emph{Swift}/XRT spectra can be fitted well only with a powerlaw with interstellar absorption, and
no additional component was required. This may be due to the further limited statistical quality of the XRT
data. The best-fit parameters of the spectral models are given in Table 3. Figure 6 shows the phase averaged
spectra for both the \emph{Suzaku} and the longest exposure XRT observations. 
The obtained spectral parameters are consistent within errors between the two \emph{Suzaku} observations. Consistency 
was also checked between the parameters obtained from the XRT and \emph{Suzaku} data by fitting the \emph{Suzaku}
spectra with the same model as for XRT. Results show that while the powerlaw photon index ($\Gamma$) and 
interstellar absorption $N_{\rm H1}$ are consistent, the powerlaw normalizations are slightly higher for the  \emph{Suzaku}
observations. The 0.3-10 \rmfamily{keV} flux obtained from all the \emph{Suzaku} and XRT observations show flux 
variations by 25 \% around the mean value of $3.5 \times \ensuremath{10^{-11}$}. Assuming the distance
to the source to be 8 Kpc, the average luminosity is in the range of $\sim$ $2-4 \times$ \ensuremath{10^{35}\,  \mathrm{erg}\,  \mathrm{s}^{-1}}.
The flux obtained from all the \emph{Swift}/XRT observations are tabulated in table 3. 
\section{Discussion}
We present a detailed timing and spectral analysis of SWJ2000.6+3210 using \emph{Suzaku} and \emph{Swift}/XRT observations. 
We established the pulse period of the source to be 890 s and investigated its pulse profile at different energy bands
from 0.3-40 \rmfamily{keV}. Though the pulse period determined is different from that reported in \cite{morris2009},
we have shown that 890 s is the strongest periodicity existing in the data and the earlier report of a periodicity at 1056 s is apparently an
artifact. The broad band 
energy spectrum can be fitted
with a powerlaw high energy cutoff and blackbody model. We did not detect any change in the spectral parameters 
for the different observations
taken with \emph{Suzaku} and \emph{Swift} ranging from November 2005 to December 2006 with the existing
quality of the data. \\

The detection of strong pulsations in the system indicate that accretion is prevalent in the source and matter 
is being channeled
along the magnetic field lines on the poles of the compact object, probably a neutron star. The nature of the companion is identified
to be a Be star through optical spectroscopic observations. Further, the X-ray pulse profiles 
and the nature of the X-ray spectra resemble that of Be X-ray binary pulsars.
But due to the lack of any certain signatures like rapid spin change, we cannot rule out 
the compact object to be a white dwarf (intermediate polar). 
The spin change rate of an accreting magnetised compact object can be expressed in terms
of the compact object parameters and the X-ray luminosity as \citep{frank}
\begin{equation*}
 \nu^{\cdot} \simeq 2.7 \times 10^{-12} \, m^{\frac{-3}{7}} \, R_{6}^{\frac{6}{7}} \, L_{37}^{\frac{6}{7}} \, \mu_{30}^{\frac{2}{7}} \, I_{45}^{-1} \, Hz \, s^{-1}
\end{equation*}
with the symbols having usual meaning. Assuming that SWJ000 has an average 0.3-40 keV X-ray flux of $15 \times10^{-11}$
  $\mathrm{erg}\,  \mathrm{cm}^{-2}\,\mathrm{s}^{-1}$ (i.e, an X-ray
luminosity of $3 \times 10^{35} \mathrm{erg}\,\mathrm{s}^{-1}$ at a distance of 8 kpc), 
the total change of pulse between the two Suzaku
observations is expected to be of the order of 2 sec for a neutron star and 0.02 sec for a white dwarf. 
However, the pulse periods measured from the two Suzaku observations are not accurate enough to
establish a sufficient change in pulse period which can rule out the possibility of a white dwarf as the
compact object. \\
A Be binary system can exhibit a wide variety of luminosities and the (0.3--70 \rmfamily{keV}) luminosity estimate of this source ( $\sim$ $2-4 \times$ $\ensuremath{10^{35}\,  \mathrm{erg}\,  \mathrm{s}^{-1}}$)
 is within the range of luminosities of this class of sources. Being an optically confirmed Be binary system, its orbital period is expected to be in the range of
100-600 days as estimated from the ''Corbet" diagram of pulse period versus orbital period \cite{corbet1984}, assuming the source to be in spin equilibrium. \\

Lastly, SWJ2000.6+3210 is characterized by a long pulse period, persistent low X-ray luminosity ($\sim \ensuremath{10^{35} \mathrm{erg}\ \mathrm{s}^{-1}}$),
low X-ray variability and probably reside in a wide binary orbit. 
All these properties indicate
that it is a persistent Be X-ray binary see \citep{reig2011}. 
The first sources discovered in this class were X Persei and RX J0146.9+6121 \citep{haberl1998}. Both of them have 
relatively low X-ray luminosities
($\sim \ensuremath{10^{34} \mathrm{erg}\ \mathrm{s}^{-1}}$) and long spin periods (837 s X Persei and 1412 s RX J0146.9+6121). Two more
sources, RX J1037.5+5647 (860 s) and RX J0440.9+4431 (202.5 s) were discovered later by \cite{reig1999} 
which belong to this class. Recently, a possible new persistent Be X-ray binary pulsar SXP 1062 was discovered in
the SMC \citep{brunet2012}. 
Incidentally, the  X-ray pulse profile of SWJ2000.6+3210 resembles that of RX J1037.5+5647 with a two peaked structure and a dip in between
both in the low and high energy bands. The pulse periods are also very similar.
It may also be noted that a very low cutoff energy (2-4 \rmfamily{keV}) was required to fit the spectra of all
these sources. We also required a low cutoff energy $\sim$ 6-7 \rmfamily{keV} to fit the energy spectra, 
in contrary to the
10-20 keV cutoff energy usually found in X-ray binary pulsars. SWJ2000.6+3210 may thus be an addition to a growing
subclass of persistent Be X-ray binaries. Additionally, the presence of a hot blackbody component
in the energy spectrum of this source further favors its classification as a Be X-ray pulsar, as all the sources
of this class discovered so far has indicated the presence of the same blackbody component 
at similar temperature and radius of the emitting
region \citep{palombara2012,palombara2013}. This hot blackbody spectral component is likely to be the neutron
star polar cap emission as indicated by its high temperature and emitting region of $< 1$ km. \\ 

A possible model for these systems is a neutron star orbiting a Be star in a wide orbit and thus accreting matter
only from the outer low density regions of the circumstellar envelope of the companion. Non detection of Fe $K\alpha$ line further supports this scenario of scanty material
in the vicinity of the neutron star.\\
It is also worthwhile to mention in this context that the claim of the source
being a heavily absorbed X-ray binary as mentioned in \cite{morris2009} is
contradictory to that found by us. Spectral fitting of the phase averaged spectra
shows that the source has only a moderate {$N_{\rm H}$} that is consistent with its optical extinction value of 4.0 \citep{masetti2008}. In addition, pulsations are also detected in the low energy range ($<$ 1
keV.)

\section{Conclusions}
The main conclusions of this work are the following :
\begin{enumerate}
 \item We have accurately determined the pulse period of the Be X-ray binary SWJ2000.6+3210 and detected
       pulsations in the source up to 40 keV.
 \item  We have probed the pulse profiles in different energy bands and established the complex
       energy dependent nature of the pulse profiles specially in the soft X-rays.
 \item We have also measured the flux of the source from observations varying over the span of a year
       and have reported a persistent low flux level varying by only a few percent. We argue that the low flux level  along with the long pulse period
and the presence of a hot blackbody spectral component indicate that
the source is a member of a growing class of persistent Be X-ray
pulsars with the other members of the class being X Per, RX J0146.9+6121, RX J1037.5-5647 and RX J0440.9+4431.
\end{enumerate}

\begin{table*} 
\scriptsize
\caption{Best fit
phase averaged spectral parameters of SWJ2000.6+3210. Errors quoted are for \textbf{90} per cent confidence range.}
\label{table}
\centering
\begin{tabular}{|c | c | c |c |c |c |c |c |}
\hline

Parameter & Obs. 1 & Obs. 2 & Sw. 1 & Sw. 2 & Sw. 3 & Sw. 4 & Sw. 5 \\
\hline
$N_{\rm H}$ ($10^{22}$ atoms $cm^{-2}$) & $1.2 \pm 0.2 $ &   $1.4_{-0.1}^{+0.2}$ & $1.2_{-0.6}^{0.7}$ & $1.7_{-0.3}^{0.4}$& -- & -- & --\\	
PowIndex & $0.84_{-0.50}^{+0.41}$  & $1.04_{-0.22}^{+0.13}$ & $0.83_{-0.37}^{0.40}$ & $1.15_{-0.21}^{0.22}$& -- & -- & --\\
Ecut (keV) & $7.0 \pm 3.8$ & $6.2_{-0.9}^{+0.8}$ & -- & -- & -- & -- & --\\
Efold (keV) & $11.2_{-3.5}^{+8.9}$ & $25.0_{-9.6}^{+25.7}$ & -- & -- & -- & -- & --\\
Norm $^a$ & $0.001_{-0.0008}^{+0.0005}$ &  $0.003 \pm 0.0009$ & $ 0.021 \pm 0.001 $ & $0.046 \pm 0.001 $ & -- & -- & --\\
$KT$  & $1.2_{-0.1}^{+0.3}$ & $1.5 \pm 0.1$& -- & -- & -- & -- & -- \\
$KT_{Norm}$ & $0.6_{-0.2}^{+0.4}$  & $0.5_{-0.1}^{+0.2}$ & -- & -- & -- & -- & --\\
 $\chi^{2}_{\nu}$/d.o.f & 1.18/343  & 1.16/358 & 0.61/21 & 0.94/60 & -- & -- & --\\
Flux (XIS) $^b$ (0.3-10 \rmfamily{keV} &  $3.2 \pm 0.01$ & $5.7 \pm 0.01$ & -- & -- & -- & -- & --\\
Flux (PIN) $^c$ (10-70 \rmfamily{keV}) &  $7.6 \pm 0.01$ & $12.3 \pm 0.01$ & -- & -- & -- & -- & --\\
Flux (XRT) $^d$ (0.3-10 \rmfamily{keV})& $3.2 \pm 0.01$ & $5.7 \pm 0.01$ & $3.4 \pm 0.01$ & $4.2 \pm 0.01$ & 3.1$\pm$ 0.01 & 3.5 $\pm$ 0.01 & 3.5 $\pm$ 0.01 \\
\hline
\end{tabular}\\
$^a$ \ensuremath{\mathrm{photons}\, \mathrm{keV}^{-1}\,\mathrm{cm}^{-2}\,\mathrm{s}^{-1}\,\mathrm{at}\, 1\, \mathrm{keV}} \\
$^b$ $^c$ $^d$  Flux is in units of \ensuremath{10^{-11}\,  \mathrm{erg}\,  \mathrm{cm}^{-2}\,\mathrm{s}^{-1}} and are in 99 \% confidence range.\\
\end{table*}

\label{lastpage}

\end{document}